\documentclass[structabstract]{aa}

\usepackage{amssymb}
\usepackage{graphicx}
\usepackage{txfonts}
\usepackage{natbib}

\begin{document}

\title{Extensions and applications of the iterative method}

\author{S.A.~Rodionov\inst{1,2,3}
\and E.~Athanassoula\inst{1}}

\institute{
Laboratoire d'Astrophysique de Marseille (LAM), UMR6110,
CNRS/Universit\'e de Provence, Technop\^ole de Marseille-Etoile, \\
38 rue Fr\'ed\'eric Joliot Curie, 13388 Marseille C\'edex 20, France
\and
Center de recherche INRIA Bordeaux -- Sub-Ouest, Bordeaux, 351
Course de la Lib\'eration, France 
\and
Sobolev Astronomical Institute, St. Petersburg State University, 
Universitetskij pr.~28, 198504 St. Petersburg, Stary Peterhof, Russia
\email{astroseger@gmail.com}
}
				  
\date{Received ????; accepted ????}


\abstract
{} 
{
We aim to develop an algorithm for constructing equilibrium initial conditions
for simulations of disk galaxies with a triaxial halo and/or a 
gaseous component.
This will pave the way for $N$-body simulations of realistic disk galaxies. 
}
{
We use the iterative method, which we presented in a previous article. The idea
of this method is very simple. It relies on constrained evolution.
}
{
We develop an algorithm for constructing equilibrium models of disk galaxies
including a gaseous disk and a triaxial or axisymmetric halo. 
We discuss two test models. The first model consists of a spherical halo, 
a stellar disk, and an isothermal gaseous disk. The second model consists of
a triaxial halo, 
a stellar disk and a star-forming gaseous disk. We demonstrate that both
test models are very close to equilibrium, as we had intended.
}
{} 

\keywords{galaxies: kinematics and dynamics -- methods: numerical}

\maketitle

\section{Introduction}

 In this article we consider a method for constructing equilibrium initial
conditions for simulations of disk galaxies with a triaxial halo and a gaseous
component. $N$-body systems
have the following feature: If we construct some
non-equilibrium $N$-body system and let it evolve, it 
will reach an equilibrium state fairly fast (on a time scale of a few
crossing-times) and this independent of the presence or absence of gas. This,
however, does not mean that it is not necessary 
to construct equilibrium initial conditions. If some
initial conditions are constructed by means of a more or less
rough approximate method, i.e. out of equilibrium, the system will readjust
fairly fast and reach some
equilibrium state. But this readjustment will change the mass
distribution of the system and its kinematics. This way an
equilibrium condition will be obtained, but it will not have the
desired parameters, i.e. the desired mass distribution and
kinematics. This makes it very difficult or in a strict
sense, nearly impossible to study a particular galaxy
with a known mass and velocity distributions. 
Also, when using $N$-body simulations to study 
an evolution as a function
of a given parameter of the initial configuration, it is necessary to
make sure that the chosen parameter does not change considerably
during the readjustment to equilibrium, or that at least the sequence
of the different models is not readjusted. Furthermore, in some cases
the initial violent readjustment to equilibrium could influence the
process under study. It is thus preferable
to construct initial conditions as close to
equilibrium as possible to avoid such problems.
 
In \citet[hereafter RAS09]{RAS09}
we presented an iterative method for constructing
equilibrium $N$-body models with given properties, the general idea of 
which can be applied to an arbitrary dynamical system. The
method outlined there can be used for constructing multicomponent
axisymmetric models of disk galaxies without gas. Here we will extend this
method for models with a gaseous disk and models with a triaxial halo.
Other methods for constructing gaseous
equilibrium disk models have been proposed. For example 
\citet{SDH05} introduced a method in which the gaseous disk
was constructed with the help of the equation of hydrostatic
equilibrium. 
 
Here, we will 
use the iterative method to construct the gaseous component
because it is conceptually simple, very
easy to implement and, as we will demonstrate in this article, gives
good results. Furthermore, in this way we are sure that all components are
in equilibrium since the iterative method has been used in all cases. 

We model the collisionless components with an $N$-body code and the gas
with SPH. We will consider two types of gas. The first type is
isothermal gas, which we model by means of the public GADGET2 code
\citep{S05}. The second type is a 
sub-resolution multiphase model for star-forming gas, developed by
\citet{SH03}.

This paper is organized as follows. We first discuss in section \ref{s_cyl} 
the application of the iterative method to the construction of 
axisymmetrical disk galaxy models with a gaseous disk. In section \ref{s_triaxial} we
demonstrate how to construct such models within triaxial haloes. We briefly
conclude in section~\ref{s_conc}.   

\section{Axisymmetric models with a gaseous disk}
\label{s_cyl}

\subsection{Equilibrium model of the gaseous disk}
\label{s_gmethod}

The vertical density profile of the gaseous disk cannot be arbitrary. It 
is governed by the balance
between gravity and pressure, i.e. by hydrostatic equilibrium:

\begin{equation}
\label{eq_h}
-\frac{1}{\rho_g} \frac{\partial P_g}{\partial z} = \frac{\partial
\Phi}{\partial z} \, ,
\end{equation}

\noindent
where $\Phi$ is the total gravitational potential of all components,
$\rho_g$ is the density of the gaseous disk and $P_g$ is the pressure
in the gaseous disk. If the equation of state has the given form 
$P=P(\rho)$, the vertical
structure of the gaseous disk is fully determined by equation
(\ref{eq_h}) for a given surface density. 
This last statement is fulfilled for both gaseous models
considered here: the isothermal gas and the sub-resolution
multiphase model \citep{SH03,SDH05}.
We thus need to construct a model of gaseous disk
with a given projected
surface density radial profile $\Sigma_g(R)$, in equilibrium in a potential
that is the sum of its own
potential and of a given external potential
$\Phi_{ext}$ generated by all other components of the galaxy.

In RAS09 we describe the iterative method that can be used for
constructing an equilibrium $N$-body system with a given mass
distribution, following given kinematical constraints. 
The same conceptually method, with only relatively minor 
modifications can be applied for the construction of the gaseous disk. It
relies on constrained (or guided) evolution. When constructing the 
equilibrium $N$-body system, we let it evolve, and during this
evolution we fix the desired mass distribution and kinematics (see
RAS09). This will be the same for the gaseous disk, except that now we
do not fix the full mass distribution, but only the projected surface
density.

\begin{figure}
\begin{center}
\includegraphics[width=8cm]{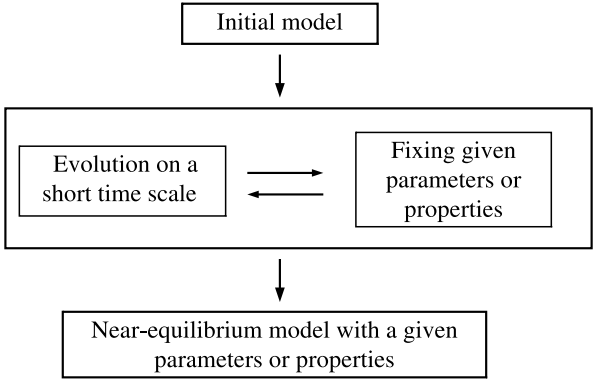}
\end{center}
\caption{General scheme of the iterative method.}
\label{fig_scheme}
\end{figure}

The general scheme of the iterative method is outlined in
Fig.~\ref{fig_scheme}. We start from some initial system, which is the
starting point of the iterative procedure. We then let the system go 
through a sequence of evolutionary steps of short duration. At the end of
each one of these steps, and before the new evolutionary step is started, 
we ``set'' the
chosen parameters or quantities to the desired values (see RAS09).
We repeat this iteration 
procedure a number of times, alternating one evolution phase and one
phase during which the necessary parameters are set, until we
come sufficiently near to the desired equilibrium state. In practice, 
to apply this general scheme we need to define which initial model we
will use, and, most important, which parameters we want to ``fix'' during 
the iterative procedure.

Let us now describe how we will apply this iterative method to the 
construction of an equilibrium gaseous disk. Our initial model is a gaseous
disk with a given surface density $\Sigma_g(R)$. 
The vertical coordinates of the particles can be arbitrarily chosen (for
example equal to zero). Equally arbitrary, the tangential velocity of 
each particle is set 
equal to the circular velocity and the vertical and radial velocities
are set equal to zero. The circular velocity is calculated from the total
potential, which is the sum of $\Phi_{ext}$, due to the adopted mass
distribution 
of the collisionless part, and the potential generated by the initial
distribution of the gas. At this stage it is not necessary
to calculate the circular velocity very accurately. Even if we set the
azimuthal velocities equal to twice or to half of the circular
velocity, the iterative method will converge to the same result. In
the case of a non-isothermal system the thermal energies of the
particles can be arbitrarily chosen. 
This model will be the starting point for the iterative procedure (see
Fig.~\ref{fig_scheme}). Having thus obtained the starting model, we start
the iterative procedure, which consists of a sequence of evolutionary steps of
short duration, followed by steps during which some parameters or quantities
are fixed, and this until a near
equilibrium model is obtained. In this example 
we fix the following parameters: 

\begin{itemize}
\item
Surface density of the gaseous disk.
\item
The condition of axisymmetry.
\end{itemize}

This means that we do not fix the vertical distribution of the particles, 
letting it adjust itself to hydrostatic equilibrium.

To fix the surface density in the gaseous disk at the end of an evolutionary
step we proceed as follows.
We construct a gaseous disk with the desired surface density, but with
velocities and vertical coordinates chosen according to the velocities
and the vertical coordinates of the disk resulting from the evolution
step.
We first construct a new gaseous disk with the desired surface 
density profile. We then
``transfer'' the velocity distribution
and the distribution of the vertical coordinates of the particles 
from the system obtained from the evolution to this new system, using the
``transfer'' algorithm described in RAS09 (see Sect. 2.2 in that
paper). The basic idea of this algorithm is very simple, namely 
we assign to the new-model particles the velocities of
those particles from the old model that are ``nearest'' to the ones in the new
model, the definition of nearest depending on the problem at hand. In
the present case we need to ``transfer'' the
vertical coordinate of the particle together with the
velocities. Since our model is axisymmetric, we need to search
for the nearest particle in the
one-dimensional space $R$, where $R$ is the cylindrical radius. This
implicitly fixes the condition of axisymmetry.
The vertical coordinate should not be taken into account during the search
for the nearest neighbour because we copy it from the evolved model particle
to the new model particle together with the velocities. In
the case of non-isothermal gas, the thermal energy of the particle should be
copied together with the velocities and the vertical coordinate. 

\subsection{Equilibrium models of stellar components}

The algorithm for constructing the equilibrium $N$-body system with
given parameters and constraints is described in RAS09 and is not
altered by the presence of the gaseous component. 
We simply need to take into account the gravitational acceleration 
caused by the gaseous component when creating the collisionless
components, since it will act on them as an external potential. 
However, because we initially know only the surface density of the
gaseous disk, we cannot calculate the three-dimensional 
gravitational acceleration. We need, therefore, to construct the
equilibrium model of the gaseous disk first, before that of the
collisionless components. After constructing this model
we have the full mass model of the gaseous disk
and can construct all other components of the galaxy,
taking into account the acceleration generated by the gaseous disk.

\subsection{Technical details}

In this section we elaborate a few important technical points, useful
to anybody wishing to apply the iteration method.

The iterative method has two free parameters: the duration of each
iteration, $t_i$, and the number of neighbours used in the
``transfer'' algorithm, $n_{nb}$ (see section \ref{s_gmethod} of this
paper and section 2.2 of RAS09). The choice of these parameters was
discussed in RAS09 section 2.5. 
We choose both these parameters empirically. 
In all experiments discussed in this article we use $n_{nb}=10$.

We construct each component of the galaxy separately in the
rigid potential of all other components. So in the iterative method when
we calculate the evolution of the system during the iteration time, we do it
in the presence of the appropriative external potential. 
This can be done either by introducing an
analytical external potential, or by adding the component(s) that create
this external potential as a rigid $N$-body system. 
In the current work we use the latter. For example, to include the
external potential due to the halo, 
we simply add rigid particles to the system  according to the mass 
distribution of the halo. 

When creating the collisionless components we follow the evolution using
the public version of the gyrfalcon $N$-body code \citep{D00,
  D02}. For the gaseous disk we  
need an appropriate SPH code. For the isothermal gas we use the public
version of the GADGET2 code \citep{S05} and for the multiphase gas code with
sub-grid resolution physics we use a private version of GADGET2
kindly provided by V. Springel and described in \cite{SH03}.

We still need to decide when the iteration procedure will be stopped. 
We can assume that the iterative process has converged when the
system does not change by more than a pre-set amount during one single
iteration. We check this convergence by comparing different
parameters of the system in the beginning and in the end of 
a single short-term evolution step, using a 
modification of the $\chi^2$ test (see Appendix \ref{s_app}). 

\subsection{Example of an axisymmetric model with isothermal gas}
\label{s_cylexample}
In this section we consider an example of a model constructed by means of the
method described above. This model is axisymmetric and has an isothermal
gaseous disk. It consists of three components: the gaseous disk, the stellar
disk, and the halo.

To start, we need to define the mass distribution in both of the
non-dissipative components and the projected surface density in the gaseous disk.
The stellar disk model is an exponential disk with a density

\begin{equation}
\label{eq_disk}
\rho_{\rm d}(R,z)=
\displaystyle\frac{M_d}{4 \pi R_d^2 z_0}
\exp\left(-\frac{R}{R_d}\right)
{\rm sech}^2\left(
\frac{z}{z_0} \right) \, ,
\end{equation}

\noindent
where $M_d$ is the total disk mass, $R_d$ is the disk scale length, $z_d$ is
its scale height and $R$ is the cylindrical radius.
The halo model is a truncated NFW halo (\citealt{N96})

\begin{equation}
\rho_{\rm h}(r) = C_{\rm h} \frac{exp(-r^2/r_{\rm th}^2)}
{(r/r_h)(1+r/r_h)^2} \, ,
\label{eq_NFW}
\end{equation}

\noindent
where $r_h$ is the halo scale length, $C_h$ is a parameter defining
the mass of the halo and $r_{\rm th}$ is the truncation radius of the halo.
Similarly to the stellar disk, the gaseous disk 
has an exponential projected surface density profile

\begin{equation}
\label{eq_gas}
\Sigma_{\rm g}(R)=
\displaystyle\frac{M_g}{2 \pi R_g^2}
\exp\left(-\frac{R}{R_g}\right) \, ,
\end{equation}

\noindent
where $M_g$ is the total mass of the gaseous disk and $R_g$ is its 
scale length.

We still need to adopt specific values for the present example. In the above we
take $M_d= 5 \cdot 10^{10} \; {\rm M}_{\odot}$, $R_d = 3 \;\rm kpc$, 
$z_0=0.6 \;\rm kpc$; $r_h=12 \;\rm kpc$, 
$C_h=0.0019 \cdot 10^{10} \rm M_{\odot}/kpc^3$, 
$r_{\rm th}=40
\;\rm kpc$; $M_g= 0.5 \cdot 10^{10} \; {\rm M}_{\odot}$, $R_g = 3
\;\rm kpc$.  
We set the temperature of the isothermal gas to $T=10000 \;\rm K$. Note that
the mass of the gaseous disk is 10\% of the stellar disk mass. 
For the chosen parameters, the total mass of the halo is 
$M_h \approx 4.9 \cdot M_d$. In this specific example we chose $N_g=100000$,
$N_d=200000$, $N_h=980311$ for
the number of particles in the gaseous disk, the stellar disk, and the halo,
respectively. With these numbers, the mass of the particles 
in the stellar disk and in the halo is the same. We use the GADGET system of
units, where the unit of length is $u_l=1 \;\rm kpc$, the unit of
velocity is $u_v=1 \;\rm km/sec$, the unit of mass is $u_m = 10^{10}
\;\rm M_{\odot}$ and consequently the unit of time is $u_t \approx
0.98 \;\rm Gyr$. For simplicity, when we convert this time unit into gigayears
we assume that $u_t=1 \;\rm Gyr$. 

 We also need to select the kinematic constrains for the stellar components 
(see RAS09). We created the disk with the following velocity dispersion profile
\begin{equation}
\label{eq_svR}
\sigma_R(R) = 100 \cdot \exp\left(-R/9\right) \; {\rm km \, s^{-1}} \, ,
\end{equation}
where $\sigma_R$ is the radial velocity dispersion.  When constructing the
halo, we did not impose any specific kinematic
constraints. Instead, we aimed for a model not far from
isotropic (see RAS09).

\begin{figure*}
\begin{center}
\includegraphics[width=14.5cm]{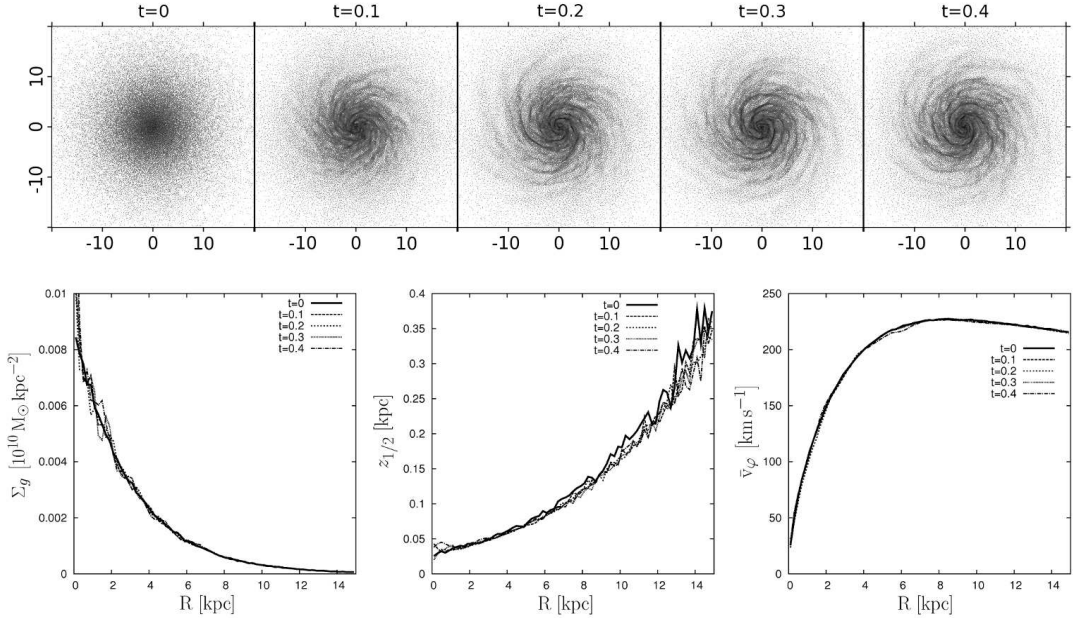}
\end{center}
\caption{Initial evolutionary stages for the gaseous disk of the constructed
disk galaxy model (axisymmetric isothermal case). 
The evolution of the model was calculated including a live halo and stellar disk.
The upper snapshots show the disk views face-on for
times 0, 0.1, 0.2, 0.3 and 0.4; the grey intensities correspond to the
logarithm of the surface density.
The bottom panels show the
dependence of various disk quantities on the cylindrical radius $R$ at
the same times. Here $\Sigma_g$ is the surface density of the gas; 
$z_{1/2}$ is the median of the $|z|$ value, i.e a
measure of the disk thickness (see \citealt{SR06}) and
$\bar v_{\varphi}$ is the mean tangential velocity.}
\label{fig_gas.iso}
\end{figure*}

 As noted in section \ref{s_gmethod}, we should first construct the
equilibrium model of the gaseous disk with the desired projected
surface density embedded in the rigid potential
generated by the halo and the stellar disk, as described in 
section~\ref{s_gmethod}.
To achieve this, we made 50 iterations, each with $t_i = 0.02 \; \rm Gyr$.
We note that $t_i$ should be shorter than the time scale of the strong
instability developing in the system under construction. In our case
the gaseous disk forms 
strong spirals relatively fast (see fig.~\ref{fig_gas.iso}). It is why we
have to choose relatively short $t_i$ in this case.

 After constructing the equilibrium gaseous disk, we have the full mass
model of the galaxy, and we can apply the algorithm for constructing
the equilibrium
models of the stellar disk and the halo (RAS09).

Let us first describe the stellar disk construction.
Our initial model was a cold disk, where all particles move on
circular orbits. We made $50$ iterations, each with $t_i=0.25 \; \rm Gyr$. The
integration step and
softening length were taken $dt=1 / 2^{10} \; \rm Gyr$ and 
$\epsilon=0.1 \;\rm kpc$,
respectively and the tolerance
parameter for gyrfalcON was set $\theta_t=0.9$. Here 
we can use relatively low precision because each iteration is short and 
errors do not accumulate (RAS09).
In order to fix the
$\sigma_R(R)$ profile we applied the
algorithm described in RAS09, section~2.3.1, with $n_{\rm div}=200$ layers. 
We used the
``transvel\_cyl'' modification of the algorithm of velocity transfer (see
RAS09). This algorithm was also used for constructing the halo.

For constructing the halo we also made $50$ iterations. The other
parameters
for this construction were  $t_i=0.5 \; \rm Gyr$, $dt=1 / 2^{10} \; \rm Gyr$, 
$\epsilon=0.1 \;\rm kpc$ and
$\theta_t=0.9$.
Our initial model was a cold model with velocities equal to zero.
During the first $10$ iterations we fixed a condition of velocity isotropy
(RAS09, 2.3.4), and we did not set any kinematic
constraints during the last $40$ iterations.

Once all three components of our model were constructed, we simply
stacked them to obtain the complete system. To check
whether this was indeed near equilibrium, as it should be, we
evolved the model using the respective GADGET2 code. 
The evolution of the
gaseous disk over 0.4 Gyr is given in Fig.~\ref{fig_gas.iso}, and 
shows that the gaseous disk conserves its structural and dynamical
properties very well.
We also checked the absence of
evolution for the stellar disk and the halo.
For reasons of conciseness we do not present the corresponding figures here.
We conclude that 
the constructed model is indeed close to equilibrium. 

\section{Models with triaxial haloes}
\label{s_triaxial}

Here we will describe how to overcome the assumption of axisymmetry and thus
to construct equilibrium models with triaxial haloes. 
Because of the powerful, yet simple basic idea of our iterative method we can
remove this assumption relatively easily. As in the axisymmetric case, 
we will construct each component of the galaxy in the potential of all other
components and then will assemble all components together to obtain
the final live equilibrium model. 
We will start from the
problem of the construction of equilibrium stellar disk in presence of the
external non-axisymmetric potential (created by all other
components of the galaxy).
 
\subsection{Construction of the stellar disk in the  
triaxial external potential}
\label{s_3disk}

The first problem whose solution is not obvious is deciding which shape
the stellar disk should have to be in equilibrium within a 
triaxial external potential. There are two
possibilities. One possibility is to somehow define the shape of the disk
and then 
try to construct an equilibrium model with this shape, i.e. model with given
mass distribution. For example, \citet{M10} obtained an approximate
shape of the disk via the epicyclic 
approximation. This, nevertheless, is only an approximation and there has been no
other, more accurate way proposed so far. For this reason, we will
let the iterative method itself find the appropriate shape of the disk.
We do it in the following way. We do not ask the iterative method to construct
a model with a fully defined space distribution of particles. Instead, we fix
only the vertical and radial distributions of particles, but not the
azimuthal distribution. This means that the $R$ and $z$ coordinates of the
particles are defined by a given distribution, but the $\varphi$ coordinate
together with velocities is not fixed.
The idea described in the last sentence is very similar to the idea
of the algorithm we used 
for constructing the gaseous disk, where we fixed only the surface
density (i.e. 
the radial and azimuthal distribution of particles) but we did not fix the 
vertical distribution of the particles (see section \ref{s_gmethod}).
Consequently the algorithm is also very similar to the ones presented
in section \ref{s_gmethod}. 

 To apply the general scheme of the iterative method (see Fig.~\ref{fig_scheme})
we need to specify an initial model and the parameters which we want to fix during
the iterative procedure. 
Our initial model is an axisymmetric disk with a given radial and vertical
distribution. The azimuthal velocity of each particle is set equal to the
circular velocity.

With this initial model, we start the iterations, which, as always,
consist of a sequence of short time evolutions followed by steps during
which we fix the desired parameters or properties (see Fig.~\ref{fig_scheme}).
In this case we fix the

\begin{itemize}
\item 
vertical and radial distribution of particles,
\item
reflection symmetry about the $XZ$ and $YZ$ planes for the case of system rotating
about the $z$-axis,
\item
given profile of $\sigma_R(R)$.
\end{itemize}

Our algorithm for fixing the vertical and the radial distribution of
particles is 
very similar to the algorithm which we used for fixing the surface density
in section~\ref{s_gmethod}. 
We construct a stellar disk with the desired vertical and radial
distribution of particles, but with
velocities and azimuthal coordinates chosen according to the velocities
and the azimuthal coordinates of the disk resulting from the evolution
step. We transfer the azimuthal coordinate of the particle 
during the ``transfer'' procedure together with the 
velocities (see section~\ref{s_gmethod}) to
the ``nearest'' particle, which in this case is found in the 
two-dimensional $R-z$ space.

There is one potential problem with the symmetry. Indeed, if we fix only
the distribution of particles in the vertical and radial directions, then 
the constructed
model can be non-symmetrical in the disk plane. Partially, this is what
we aimed for because our target was to avoid axisymmetry. 
However, even in a triaxial potential we would like the equilibrium model to
have a certain level of symmetry and a smooth density distribution. We can
enforce 
our model to have reflection symmetry about the $XZ$ and $YZ$ planes by fixing
these symmetries during the iterative procedure (see Fig.~\ref{fig_scheme}).
But we should take into account that our disk rotates about the $z$-axis. 
Reflection symmetry about the $XZ$ plane is fixed by changing 
the sign of $y$ and $v_x$ for all particles with a probability of
$1/2$. Reflection symmetry about 
the $YZ$ plane is fixed by changing the sign of $x$
and $v_y$ with a probability of $1/2$. 
 
Let us note, however, that our experiments show for the
stellar disks that fixing this symmetry is not so crucial.
In most cases, stellar disks constructed with or without this
symmetry are quite similar. Nevertheless, in a few cases 
disks constructed without imposing this symmetry can be slightly clumpy
at the periphery.

Similar to the axisymmetric case, we construct a model with a given $\sigma_R(R)$
profile. The value of $\sigma_R$ is defined as 
the radial velocity dispersion of all particles in a cylindrical layer.
To fix this value we use the algorithm described in section~2.3.1 of RAS09.
However, we note that the physical interpretation of the $\sigma_R(R)$
profile in a triaxial case is not so straightforward. 

\subsection{Construction of the gaseous disk in the triaxial external
potential} 
\label{s_trigas}

 Here we need to unite the approach discussed in the previous section with the
method described in section~\ref{s_gmethod}.
We should allow the gaseous disk
more freedom here than in the axisymmetric case, and thus will constrain
only the radial distribution of particles (i.e. the radial density profile).

Our initial model is the same as in the axisymmetric case (see Sect.~\ref{s_gmethod}).
During the main part of the iterative procedure (see Fig.~\ref{fig_scheme})
we fix the following parameters:

\begin{itemize}
\item 
Radial distribution of particles in the gaseous disk.
\item
Reflection symmetry about $XZ$ and $YZ$ planes for the case of a
system rotating about the $z$-axis.
\end{itemize}

In section~\ref{s_gmethod} we described the algorithm for fixing the surface
density of the gaseous disk. To fix only the radial distribution of particles we need
to make minor modification to this algorithm. We need to transfer the azimuthal
coordinate of each particle, together with their velocities and vertical
coordinate (and thermal energy in the case of non-isothermal gas).
So we construct a gaseous
disk with the desired radial density profile, but with velocities as well as
vertical and azimuthal coordinates taken from the disk resulting from the short
evolution step. 

The algorithm for fixing reflection symmetries was discussed in the previous 
subsection. We note, however, that in some cases this algorithm of construction
non-axisymmetric gaseous disk may not work very well. The problem is that
a reflection symmetry about the $XZ$ and $YZ$ planes does not necessarily
imply that the model has  
a smooth density distribution and for some equations of state the constructed
model can have a clumpy density distribution. This can well
happen in the case of gas which tends to form clumps very fast.
We met this problem
when we tried to construct isothermal gaseous disks for
temperature of $10000$ K. .
In this case one can use the old
algorithm (section~\ref{s_gmethod}) and construct an axisymmetrical gaseous
disk. Such a disk cannot be
in equilibrium in a triaxial external potential,
it may, however, be not so far from equilibrium and have a smooth density
distribution. 

\subsection{Construction of the triaxial halo}
 We need to construct a triaxial halo that should have a given mass
distribution and be in equilibrium in a potential consisting of the halo
self-potential and the external potential generated by all other components
of the galaxy. 

The initial model for our iterative method is the halo with the given mass
distribution and with arbitrary, e.g., zero velocities.
During the main part of the iterative procedure (see Fig.~\ref{fig_scheme})
we fix the following parameters:

\begin{itemize}
\item
The given mass distribution.
\item
Reflection symmetry about the $XZ$ and $YZ$ planes for a non-rotating 
system (see discussion below).
\item
During the first 10 iterations we fix a condition of velocity isotropy (see
discussion below). 
\end{itemize}

 In some cases the model we constructed fixing only the 
mass distribution had a little rotation. If we wish to
prevent this, we need to fix the
condition of non-rotation during the iterative procedure. There are several
ways to do this, and we used the following one.
We fixed a condition of reflection symmetry about the $XZ$ and
$YZ$ planes for the case of non-rotating system, which, however, is 
different
from the reflecting symmetry which we used in
section~\ref{s_3disk} for the rotating system. In the present case, the
reflected symmetry about the $XZ$ plane is achieved by
changing the sign of $y$ and $v_y$ for all particles with a probability of 
$1/2$. Similarly, the reflection symmetry about the $YZ$ plane is achieved 
by changing the sign of $x$ 
and $v_x$ with a probability of $1/2$. It is easy to show that after 
this symmetrization the
model will not have net rotation about any of three axis.

 During the first 10 iterations we fix the condition of velocity isotropy.
We do it to avoid constructing a model with high velocity anisotropy 
(see discussion in RAS09 section~3.1).

\subsection{Construction of the multicomponent system}
\label{s_trimult}

 We will now construct the multicomponent equilibrium model consisting
of the triaxial halo and the stellar and gaseous disks. However, we again have 
the problem that
we initially do not know the full three-dimensional mass distribution of
the stellar and gaseous disks. We construct each component in the external
potential generated by all other components of the galaxy. 
To construct the gaseous disk we need the final
mass distribution in the stellar disk, which we do not have initially, 
and, moreover,
to construct the stellar disk we need the final density distribution in
the gaseous disk. However, in both
cases (for gaseous and for stellar disks) we fix the 
radial mass distribution so that the initial axisymmetric disk and the final 
non-axisymmetric disk have the same
radial mass profile. Therefore, the acceleration generated by
the initial axisymmetric disk is not so very different from the
acceleration generated by the final non-axisymmetric disk.
Consequently, even if we construct a stellar disk that is in
equilibrium with the initial 
axisymmetric gaseous disk (and of course with the halo), 
this disk will be practically in equilibrium with the final
non-axisymmetric gaseous disk. But to be sure, we use the following simple
two step procedure.
\begin{enumerate}
\item
Construction of an equilibrium stellar disk in the presence of the initial
axisymmetric gaseous disk and halo. Let us call this model D1.
\item
Construction of gaseous disk G1 in the presence of D1 and halo.
\item
Construction of stellar disk D2 in the presence of G1 and halo.
\item 
Construction of gaseous disk G2 in the presence of D2 and halo.
\item
Construction of the halo in the presence of D2 and G2.
\end{enumerate}

 The final model consists of the constructed halo, the stellar disk D2, and
the gaseous component G2.

\subsection{Example of the model}
\label{s_etri}
In this section we discuss the construction of a model of a disk galaxy
with a triaxial halo with an axial ratio 1:0.75:0.5 and
with a gaseous disk modelled by means of the sub-resolution multiphase model 
for star-forming gas of \citet{SH03}. 
The parameters of the gas that control star formation and feedback were taken
from \citet{SH03}, where these parameters were chosen
to fit Kennicut's law \citep{K98}.

The mass model of the triaxial halo (i.e. the initial model for our iterative
procedure) is constructed in the following way.
We take a spherical halo and squeeze 
it along the $Y$ and $Z$ axes by multiplying the $y$-coordinate and
$z$-coordinate of each particle by $b=0.75$ and $c=0.5$, respectively.
The initially spherical haloes have the density profile described in
\citet{H93}, namely

\begin{equation}
\rho_{h}(r) = \frac{M_{h}}{2 \pi^{3/2}} \frac{\alpha}{r_{c}}
\frac{\exp{(-r^{2}/r_{c}^{2})}}{r^{2}+\gamma^{2}},
\end{equation}
where $M_h$ is the mass of the halo, $\gamma$ is the core radius and $r_{c}$ is the
cutoff radius.
The normalization constant $\alpha$ is defined by
\begin{equation}
\alpha = \{ 1 - \sqrt{\pi} q \exp{(q^{2})} [1- \textrm{erf}(q)] \}^{-1}  \, ,
\end{equation}
where $q=\gamma / r_{c}$ \citep{H93}. 
In the model presented here $M_h = 25 \cdot 10^{10} \; \rm M_{\odot}$, 
$\gamma = 1.5 \; \rm kpc$ and $r_c = 30 \; \rm kpc$.

We obtain the initial mass model for the stellar disk as in the previous axisymmetric 
example. The initial model of the gaseous disk has the surface density profile
(\ref{eq_gas}) with parameters $R_g = 3 \; \rm kpc$ and 
$M_g = 1 \cdot 10^{10} \; \rm M_{\odot}$, i.e. this model has a  gaseous
component twice as
massive as the previous one. The number of particles for
the gaseous and stellar disks and for the halo are $N_g=200000$, $N_d=200000$,
$N_h=1000000$, respectively. We note that initially the stellar and gaseous disks are
axisymmetric, 
but during the iterative procedure the azimuthal distribution
of particles can be changed, so disks can become non-axisymmetric.

Following the procedure described in Sect.~\ref{s_trimult}, we make $100$ 
iterations to construct the stellar disk D1, $50$ iterations to 
construct the gaseous disk G1, $10$
iterations to construct D2 and G2, and $50$ iterations for
the halo. All other parameters of the iterative procedure are the same as in
the previous example. But we note that to calculate
the evolution of the gaseous disk during the iterative process we use
the corresponding GADGET2 with star formation and feedback.

\begin{figure*}
\begin{center}
\includegraphics[width=14.5cm]{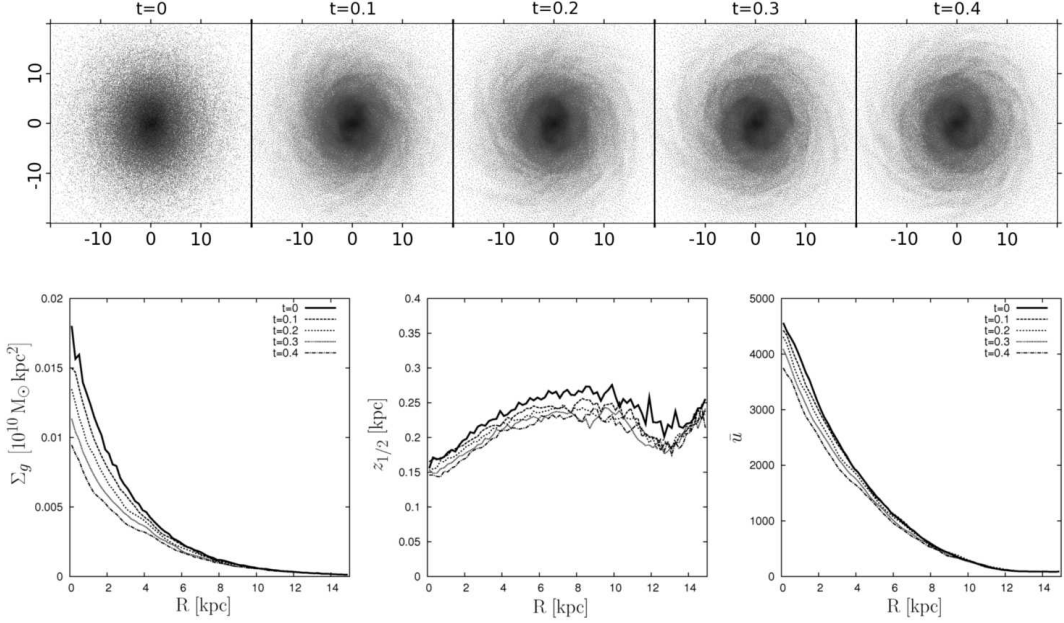}
\end{center}
\caption{Initial evolutionary stages for the gaseous disk of the constructed
disk galaxy model (with triaxial halo and star-forming gas). 
We show the same quantities as in Fig.~\ref{fig_gas.iso}, 
except for the bottom right panel, which shows
the dependence of the mean thermal energy on $R$.}
\label{fig_gas.tri}
\end{figure*}

\begin{figure*}
\begin{center}
\includegraphics[width=14.5cm]{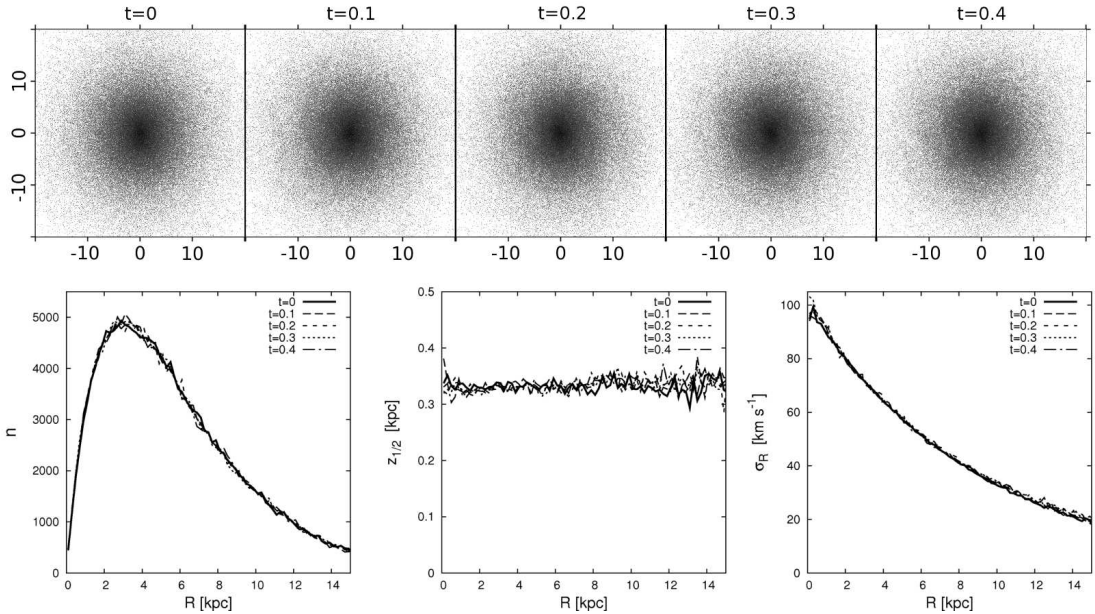}
\end{center}
\caption{
Initial evolutionary stages for the stellar disk of the constructed
disk galaxy model (with triaxial halo and star-forming gas). 
The upper snapshots show the disk viewed face-on for
times 0, 0.1, 0.2, 0.3 and 0.4; the grey intensities correspond to the
logarithms of particle number density.
The bottom panels show the
dependence of various disk quantities on the cylindrical radius $R$ at
the same times.
Here $n$ is the number of particles in concentric
cylindrical layers; $z_{1/2}$ is the median of the value $|z|$, and
$\sigma_R$ is the radial velocity dispersion.}
\label{fig_disk.tri}
\end{figure*}

\begin{figure*}
\begin{center}
\includegraphics[width=5cm, angle=-90]{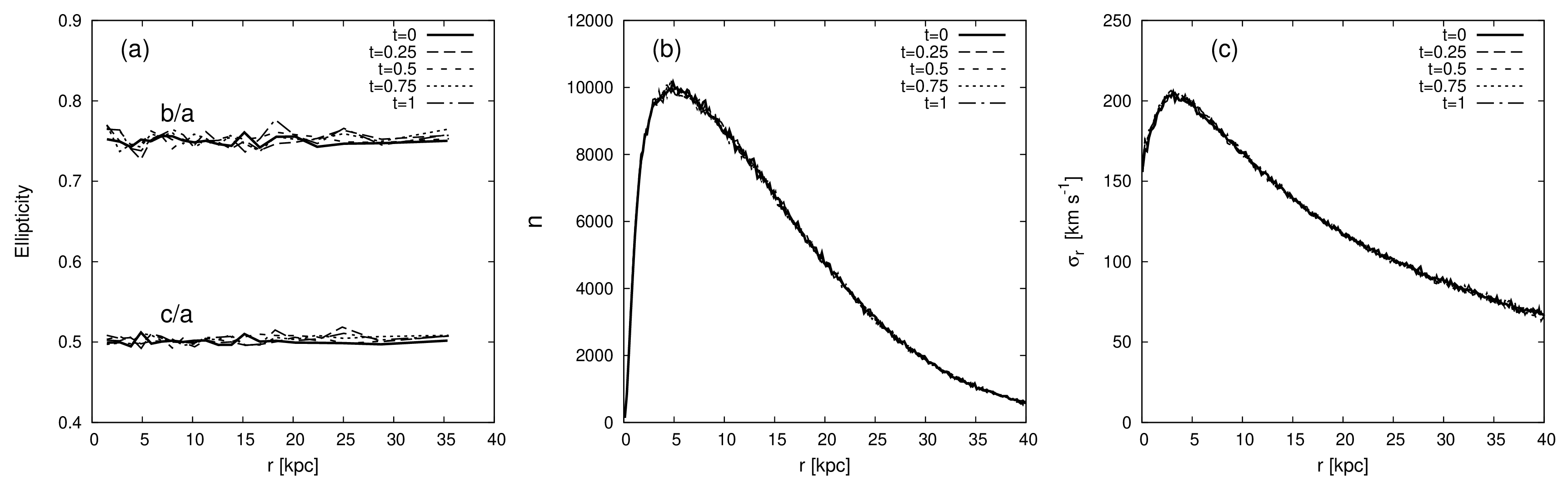}
\end{center}
\caption{Initial evolutionary stages for the halo of the constructed
disk galaxy model (with triaxial halo and star-forming gas). 
We show the dependence of various quantities on the spherical radius $r$ for
various moments of time. (a) shows radial profiles of axial ratios $b/a$ 
and $c/a$. These
profiles were calculated by means of the method described in \citet{M10}.
(b) shows the profile of the number of particles $n$ in
concentric spherical layers. (c) shows the profile of the radial 
velocity dispersion.}
\label{fig_halo.tri}
\end{figure*}

As in the previous case, we check the equilibrium of the constructed model.
The evolution of the gaseous disk over $0.4$~Gyr is shown 
in Fig.~\ref{fig_gas.tri}.
It is clear that the surface density of the gaseous disk is gradually 
decreasing
because of star-formation. At the same time, the thickness of the gaseous disk and
the mean thermal energy are also gradually diminished. But this evolution of
the gaseous disk is fairly slow. On Fig.~\ref{fig_disk.tri} we show the
evolution of the stellar disk on a time scale of $0.4$~Gyr. It demonstrates
that the stellar disk is very close to equilibrium. There is absolutely no
sign of initial adjustment to equilibrium. But we note that this stellar
disk will form a bar after $2$~Gyr of evolution. Fig.~\ref{fig_halo.tri} shows
that the halo did not change its properties during the first gigayear of the
evolution, and that it perfectly kept its triaxial shape. 
We can conclude that the model constructed in this way is indeed very close to 
equilibrium, as it should be.

\subsection{Two-arms spiral in a gaseous disk in a tri-axial potential}

\begin{figure*}
\begin{center}
\includegraphics[width=15cm]{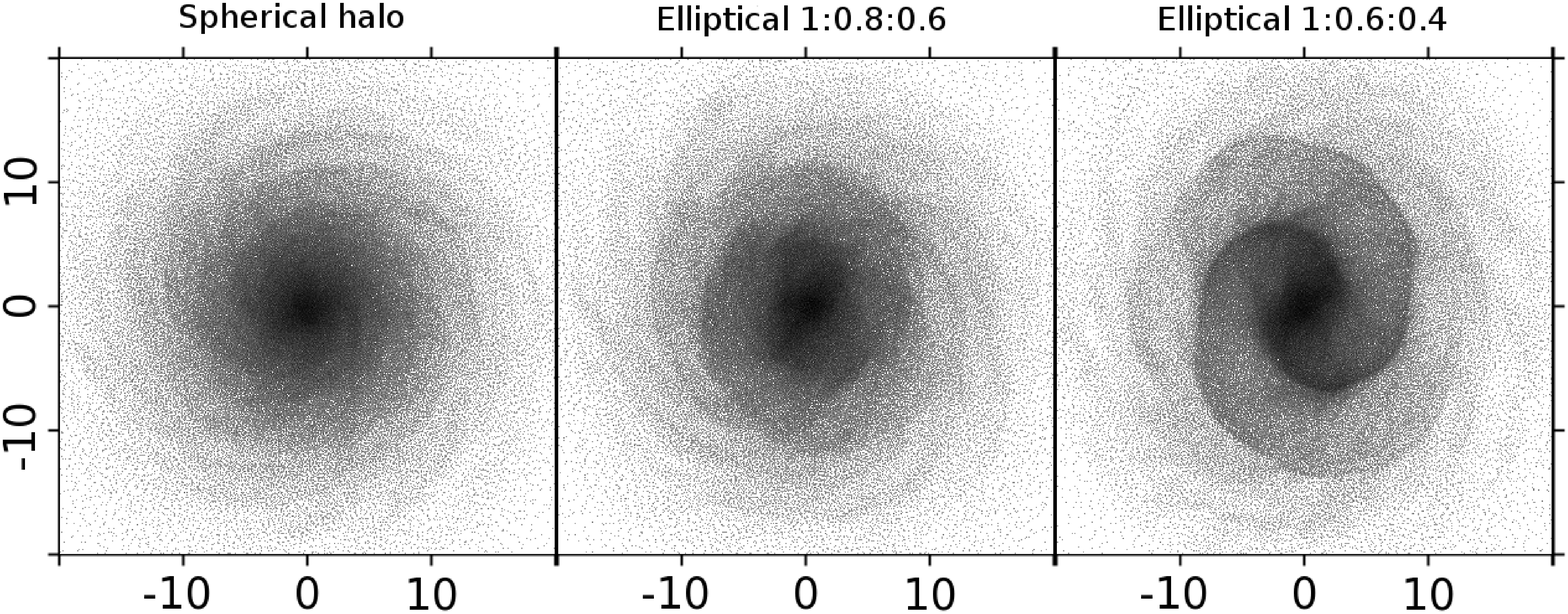}
\end{center}
\caption{Gaseous disk from three models with different holes
at time $0.5$ Gyr. From left to right: spherical
halo, tri-axial halo with $a:b:c=1:0.8:0.6$, and tri-axial halo with
$a:b:c=1:0.6:0.4$. These models are similar to the one
described in section \ref{s_etri}.} 
\label{fig_gas.gtr}
\end{figure*}

Comparing the initial evolution of the gaseous disk in the two examples
presented in Figs.~\ref{fig_gas.iso}~and~\ref{fig_gas.tri}, we find
very pronounced differences. In particular,
the isothermal gas forms a more fine-grained
structure, which can be explained because it is much colder than our
star-forming gaseous disk. On the other hand, our second model, which
has a triaxial halo and a multiphase gas, shows a fairly pronounced,
grand design, two-armed spiral. Because there is more than one
difference between these two models, concerning both the gas model and
the halo radial profile and shape, it 
is not obvious from the onset to which one the differences in the spiral
structure are due. To elucidate this we ran a few more models and saw
that grand-design two-armed spirals are found in all  
models with sufficiently triaxial halos. Figure~\ref{fig_gas.gtr} 
shows the face-on
view of three gaseous disks at $t$ $0.5$ Gyr. These come from three
models that have the same multi-phase gas description, the same halo
radial density profile, but different halo shapes. The model with
the spherical halo (left-most panel) shows a multi-arm spiral, although less
fine-grained than the isothermal gas model described in section 2. On
the other hand, the model with the most triaxial halo has a strong
two-armed, grand design spiral and the model with a weakly triaxial
halo a much less strong spiral. I.e., there is a clear tendency     
that the stronger the ellipticity, the more pronounced the two-armed
spiral structure will be. Our simulations also show that, like the halo,
this spiral 
structure does not rotate and always keeps the same orientation  
(see figs.~\ref{fig_gas.tri}~and~\ref{fig_gas.gtr}). The above discussion
shows that this two-armed structure is connected to the halo triaxiality.
It survives until the time when the
bar starts to grow in the stellar disk. We therefore believe that 
this two-armed structure is caused by the forcing of the triaxial halo.

\section{Conclusions}
\label{s_conc}

 In this paper we successfully applied the iterative method presented in our 
previous article (RAS09) to a particularly difficult problem, namely the 
construction of equilibrium initial conditions for
simulations of disk galaxies including a gaseous component and/or a 
triaxial halo.
Our iterative method relies on constrained (or guided) evolution, 
and is both
simple and powerful. In our previous article we presented an algorithm for
constructing an axisymmetric multicomponent model of a disk galaxy. 
Here we extended the 
application to models with gaseous
components and models with non-axisymmetric haloes.
We thus developed an algorithm for constructing an equilibrium
gaseous disk in the presence of an axisymmetric or non-axisymmetric external
potential. We tested our method for two types of gas. The first type is
isothermal gas, and the second type is a sub-resolution multiphase model for
star-forming gas, developed by \citet{SH03}. We also presented an algorithm for
constructing a stellar disk embedded in a non-axisymmetric external potential.

 We discussed two test models of disk galaxies. The first model consists of
a spherical halo, a stellar disk, and an isothermal gaseous disk. 
The second model consists of a triaxial halo, a stellar disk, and a 
star-forming gaseous disk. We
demonstrated that both models are very close to equilibrium. In all cases,
we gave sufficient explanations to allow the reader to reproduce the
algorithm in his/her own code. 

Although algorithms for creating isolated triaxial systems (e.g. for
elliptical galaxy models) have already been presented elsewhere
\citep{S79, RAS09, D09}, this is, to our knowledge, the first proposed
algorithm for a multicomponent disk galaxy system, with a triaxial halo and
non-axisymmetric gaseous and stellar disks. Thus this work paves the way to
many studies of realistic galaxy disks and the formation and
evolution of their structures and substructures. 

\section*{Acknowledgements}
We thank V. Springel for the use of a non-public version of his
GADGET2 code,
which includes sub-grid physics and the referee, C. Struck, for a
constructive report.
This work was partially supported by grant ANR-06-BLAN-0172,
by the Russian Foundation for
Basic Research (grants 08-02-00361-a and 09-02-00968-a)
and by a grant from the President of the Russian
Federation for support of Leading Scientific Schools (grant NSh-1318.2008.2).

\appendix

\section{Checking the convergence}
\label{s_app}

In simulation projects on galaxies it is customary to run a relatively
large number of simulations to inter-compare and understand the effect
of various parameters. Thus, creating the initial conditions can be a
considerable part of the work and it makes sense to streamline
it. In particular, the amount of CPU involved depends on the number of
iteration steps made. This number should be sufficiently large, so that the
iteration procedure can converge, but not excessively large, so as
not to needlessly waste time. It is thus necessary to be able to assess
whether
the iteration has converged or not. The most straightforward way is
of course to plot the evolution in time of various radial profiles
(such as the density, the mean velocities and dispersions etc) and
check by eye whether the variation between the two last iteration times is
sufficiently small. This, however, can be very tedious, particularly
if it is carried out a number of times for each initial conditions. It
is thus useful to prepare tools that can give information on whether a
rough convergence has been achieved, before starting the visual
examination. In this appendix we will describe how this can be carried
out in practice. 
We aim to compare the system in the beginning and in the end of
a short-term evolution during a single iterative step.
So we need tools to compare
two $N$-body models (in this context, a gaseous disk consisting of SPH
particles
can also be considered as an $N$-body model). We note that
the following algorithm is fairly similar to a test of the statistical
hypothesis 
that two $N$-body systems are just two random realizations of the same
distribution function (hereafter DF). Our algorithm is based on comparisons
of profiles of different quantities.  
 
We wish to compare profiles of some quantity $Q$
 along some axis $A$ for both systems. We divide these systems into pieces
along 
the axis $A$ in a way that each piece contains approximately the same 
number of particles. For this, 
we divide the first system into pieces, each
containing the same number of particles and calculate the corresponding 
boundaries of these 
pieces in the first system. We then divide the second system by means of
these boundaries. In each piece we calculate a given quantity whose
value we denonte by $Q$. Let $q_{1,i}$,
$q_{2,i}$ be the calculated values in the $i$-th piece in the first and 
the second model, respectively. If we consider the $N$-body system as 
a random realization of some DF, then each value calculated in
this system can be considered as a random variable. Let $Q_i$ be random
variable defined as the value $Q$ calculated in piece $i$.
For example, if the two
systems under consideration are just two random realization of the same
DF, then $q_{1,i}$ and $q_{2,i}$ are two samples of the random variable $Q_i$.
We can estimate the variance of this random variable. 
Let $var_{1,i}$ and $var_{2,i}$ be estimates 
of the variance of $Q_i$ calculated for the first and the second system,
respectively, and let us consider the value
\begin{equation}
\label{eq_c}
C = \sum_{i=1}^{n} \frac{(q_{1,i} - q_{2,i})^2}{var_{1,i} + var_{2,i}} \, ,
\end{equation}
where $n$ is the total number of pieces. If the distribution of $Q_i$ is 
close to a normal distribution 
and $var_{1,i}$ and $var_{2,i}$ are close to the real variance, then the
distribution of $C$ should be
close to a chi-square distribution with $n$ degrees of
freedom $\chi^2_n$ \citep{K51}. Such a distribution has a mean equal to $n$
and a variance 
equal to $2\,n$. According to the central limit theorem, as $n$ tends to
infinity, the $\chi^2_n$ tends to normal distribution. 
So in a first approximation the values of 
\begin{equation}
H = \frac{C - n}{\sqrt{2\,n}} \, ,
\end{equation}
have a distribution close to the normal
distribution with mean equal to 0 and variance equal to 1. Consequently, if 
value $H \lesssim 3$, then the value of $C$ can be explained by the fact that
the number of particles is finite, even in the case when the two $N$-body systems under consideration
are just two random realizations of the same DF. 
Otherwise, if $H \gtrsim 3$, then
these $N$-body systems probably differ, 
because the value of $C$ can be hardly explained by the fact that
the number of particles is finite.
In other words, the value of $H$ gives us the distance between the two
profiles of the quantity Q (calculated for the two different $N$-body systems)
measured in sigmas.

 We use three types of quantities $Q$: the number of particles in the piece, 
the mean of some quantity, and the dispersion of some quantity. We
need to estimate the variance of the random variable $Q_i$ in these cases. Let us
choose some parameter of the particles $f$. If $Q$ is the mean of the values
of $f$, then the variance of the random variable $Q_i$ (the variance of
the sample mean)  
can be estimated as 
\begin{equation}
var_m = \sigma_{f,i}^2/n_i \;,
\end{equation}
where $\sigma_{f,i}^2$ is the variance of the quantity $f$ and $n_i$ is 
the number 
of particles in piece $i$. If $Q$ is the variance of the values of $f$ then the
variance of random variable $Q_i$ (this is the variance of sample variance) 
can be calculated as
\begin{equation}
var_{var} = \frac{(n_i-1)^2}{n_i^3} m_4 + \frac{(n_i-1)(n_i-3)}{n_i^3} m_2^2 \; ,
\end{equation}
where $m_2=\sigma_{f,i}^2$ and $m_4$ are the second and fourth central moments 
of variable $f$ in piece $i$ \citep[p. 164]{K51}.
And in the last case then $Q$ is the number of
particles in piece $i$. For a fixed division of the system the
random variable $Q_i$ has a binomial distribution. And if 
$n_i \ll N$, where $N$ is total number of particles, the variance of $Q_i$
can be estimated as $var_n = n_i$. But we note that in our case 
the situation is 
more complicated because our algorithm of division of the systems. But we
assume that this formula is valid in our case, more precisely,
we assume that in
equation (\ref{eq_c}) $var_{1,i} + var_{2,i} = n_{1,i} + n_{2,i}$.

To check the convergence of the iteration in terms of some value
$Q$, we calculate the value of $H$ for the last 10 iterations. If the mean
of these ten values $H_{10}$ is less than $3$, then it means that profile of
$Q$ does not change at all during one iteration. It is a very strict criterion,
and for practical purposes we assume that iteration has converged if
$H_{10}$ is less than $10$.
\end{document}